\documentclass[
12pt,
preprint,preprintnumbers,nofootinbib,
groupedaddress,superscriptaddress,amsmath,amssymb]{revtex4}
\usepackage{graphicx}
\usepackage{dcolumn}
\usepackage{bm}
\usepackage{amssymb}
\usepackage{amsmath}
\usepackage{epsfig}    
\usepackage{color}
\usepackage{slashed}
\usepackage{hhline}

\def\be{\begin{equation}}
\def\ee{\end{equation}}
\newcommand{\bea}{\begin{eqnarray}}
\newcommand{\eea}{\end{eqnarray}}
\newcommand{\nn}{\nonumber}

\numberwithin{equation}{section}

\begin{document}

{\begin{flushright}{KIAS-P18024}
\end{flushright}}

\title{Zee-Babu type model with $U(1)_{L_\mu - L_\tau}$ gauge symmetry}
%

\author{Takaaki Nomura}
\email{nomura@kias.re.kr}
\affiliation{School of Physics, KIAS, Seoul 02455, Korea}

\author{Hiroshi Okada}
\email{macokada3hiroshi@cts.nthu.edu.tw}
\affiliation{Physics Division, National Center for Theoretical Sciences, Hsinchu, Taiwan 300}

\date{\today}

\begin{abstract}
We extend the Zee-Babu model introducing local $U(1)_{L_\mu - L_\tau}$ symmetry with several singly-charged bosons.  
 We find a predictive neutrino mass texture  in a simple hypothesis that mixings among singly-charged bosons are negligible.
 Also lepton flavor violations are less constrained compared with the original model.
 Then we explore  testability of the model focussing on a doubly-charged boson physics at the LHC and the ILC.
\end{abstract}
\maketitle
\newpage

\section{Introduction}
Radiative seesaw models are one of the promising candidates not only to establish neutrino mass matrix radiatively but also to have high testability for new physics at current and future experiments.
Zee-Babu model is the minimal scenario that does not require any additional fermions but only two charged bosons; singly-charged one ($h^\pm$) and doubly-charged one ($k^{\pm\pm}$)~\cite{2-lp-zB}, where the neutrino mass matrix is arisen at the two loop level.
 It is found a phenomenological prediction on neutrinos; a massless active neutrino is predicted due to antisymmetry of the neutrino mass structure. The serious analyses are found in refs.~\cite{Herrero-Garcia:2014hfa, AristizabalSierra:2006gb}, considering lepton flavor violations (LFVs).~\footnote{There are several variation models applying Zee-Babu model~\cite{Okada:2014qsa, Nomura:2016rjf, Nomura:2016ask, Guo:2017gxp, Okada:2014vla, Araki:2010kq, Bonilla:2016diq, Cao:2017xgk, Baek:2012ub}.}
Also one could find new signals of new bosons,
especially, the mass of $k^{\pm\pm}$ and its related interacting couplings can be constrained by the LEP experiment~\cite{Schael:2013ita} and can be well-measured by current LHC experiments~\cite{ATLAS:2017iqw,Aaboud:2017qph} and by future international linear collider (ILC) experiment~\cite{Baer:2013cma, Fujii:2017vwa} applying the chiral polarizations of beam, which might be distinguished from those in the other models~\cite{Nomura:2017abh}. 

In this letter, we extend the original Zee-Babu model by imposing a gauged $U(1)_{L_\mu-L_\tau}$ symmetry with several singlet bosons having $L_\mu - L_\tau$ charge where the Yukawa couplings associated with neutrino mass generation are constrained by the symmetry realizing predictable mass strructure
\footnote{This gauge symmetry with different types of radiative seesaw models are found in refs.~\cite{Baek:2015mna, Lee:2017ekw}.}.
 In addition, this gauge symmetry has phenomenologically interesting properties; gauge anomaly is canceled~\cite{He:1990pn,Foot:1994vd}, excess of muon anomalous magnetic dipole moment (muon $g-2$) can be explained~\cite{Gninenko:2001hx,Gninenko:2014pea,Altmannshofer:2016brv}, lepton flavor non-universality in semileptonic $B$-meson decays can be addressed with some extensions~\cite{Altmannshofer:2014cfa,Crivellin:2015mga,Altmannshofer:2016jzy,Ko:2017yrd,Chen:2017usq}, and other interesting studies can be found in~\cite{Heeck:2014qea,Baek:2015fea,Heeck:2016xkh,Altmannshofer:2016oaq,Patra:2016shz,Biswas:2016yan,Heeck:2011wj, Nomura:2018yej,Chen:2017gvf,Asai:2017ryy,Kaneta:2016uyt,Araki:2017wyg,Chen:2017cic}.
In our analysis we find that restricted Yukawa couplings coming from the additional symmetry lead to a predictive neutrino texture in a simple hypothesis that mixings among singly-charged bosons are negligibly tiny.
Furthermore LFV constrains are
much relaxed compared to the original model in such a small mixing scenario. Then we will focus on analyzing doubly-charged boson at the collider experiments such as the LHC and the future ILC, and discuss testability of our model taking into account current constraints at the LHC and the LEP.

This paper is organized as follows.
In Sec.~II, we show our model, and formulate neutrino mass matrix.
In Sec.~III, we analyze the doubly-charged boson at colliders and show results.
Finally We conclude and discuss in Sec.~IV.


 \begin{widetext}
\begin{center} 
\begin{table}[t]
\begin{tabular}{|c||c|c|c|c|c|c||c|c|c|c|c|c|}\hline\hline  
Fields & ~$H$~ & ~$h^+_{-1}$~ & ~$h^+_{0}$~& ~$h^+_{+1}$~ &~$k^{++}$~ & ~$\varphi$~ & ~$L_{e}$~ & ~$L_{\mu}$~ & ~$L_{\tau}$~ & ~$e_{R}$~ & ~$\mu_{R}$~ & ~$\tau_{R}$~ 
\\\hline 
 $SU(2)_L$ & $\bm{2}$  & $\bm{1}$  & $\bm{1}$ & $\bm{1}$ & $\bm{1}$  & $\bm{1}$ & $\bm{2}$& $\bm{2}$& $\bm{2}$ & $\bm{1}$& $\bm{1}$& $\bm{1}$   \\\hline 
$U(1)_Y$ & $\frac12$ & $1$  & $1$ & $1$  & $2$ & $0$ & $-\frac12$ & $-\frac12$ & $-\frac12$ & $-1$ & $-1$ & $-1$    \\\hline
 $U(1)_{L_\mu-L_\tau}$ & $0$ & $-1$  & $0$ & $1$   & $0$   & $1$  & $0$   & $1$  & $-1$ & $0$   & $1$  & $-1$    \\\hline
\end{tabular}
\caption{Field contents of bosons and fermions in the lepton sector
and their charge assignments under $SU(2)_L\times U(1)_Y\times U(1)_{L_\mu - L_\tau}$.}
\label{tab:1}
\end{table}
\end{center}
\end{widetext}

\section{A model}

In this section we introduce our model in which neutrino masses are generated at two-loop level and $U(1)_{L_\mu - L_\tau}$ gauge symmetry is imposed.
Fermion sector is the same as the SM one where leptons have $U(1)_{L_\mu - L_\tau}$ charge as shown In Table~\ref{tab:1}.
In scalar sector, we introduce three singly charged scalar and one doubly charged scalar fields which are $SU(2)$ singlet; singly charged scalar fields have 
$U(1)_{L_\mu - L_\tau}$ charge $+1$, $0$ and $-1$ while doubly charged scalar field does not have $L_\mu -L_\tau$ charge.
Here we write singly charged scalars as $h^+_{Q_{L_\mu -L_\tau}}$ with electric ($L_\mu -L_\tau$) charge $+1(Q_{L_\mu -L_\tau})$ and complex conjugate is defined as 
$(h^+_{Q_{L_\mu -L_\tau}})^* = h^{-}_{-Q_{L_\mu -L_\tau}}$.
In addition we introduce SM singlet scalar $\varphi$ with $L_\mu -L_\tau$ charge to break $U(1)_{L_\mu - L_\tau}$ gauge symmetry and to give mass to $Z'$ boson from the new $U(1)$.
The SM Higgs $H$ and $\varphi$ develop VEVs and we write these fields as 
\begin{equation}
H =\left[\begin{array}{c} w^+\\ \frac{v_H + \tilde{h} +i z}{\sqrt2} \end{array}\right],\ \varphi = \frac{1}{\sqrt{2}} (v_\varphi + \varphi_R + i z')
\end{equation}
where $w^+$, $z$ and $z'$ are Nambu-Goldstone(NG) bosons absorbed by $W^+$, $Z$ and $Z'$.

The Yukawa couplings associated with charged scalar fields are given by
\begin{align}
L_Y = & f_{e \mu} \bar L^c_{L_e} (i \sigma_2) L_{L_\mu} h_{-1}^+ + f_{\mu \tau} \bar L^c_{L_\mu} (i \sigma_2) L_{L_\tau} h_{0}^+ + f_{e \tau} \bar L^c_{L_e} (i \sigma_2) L_{L_\tau} h_{+1}^+ \nn \\
& + g_{ee} \bar e^c_R e_R k^{++} + g_{\mu \tau} \bar \mu^c_R \tau_R k^{++} + h.c. \, ,
\end{align}
where $\sigma_2$ is the second Pauli matrix.
Note that the coupling $f_{ab}$ is anti-symmetric due to nature of anti-symmetry under $SU(2)_L$ indices in the corresponding operators~\cite{2-lp-zB}
\footnote{ The flavor-diagonal term of $\bar L^c_{L_e} (i \sigma_2) L_{L_e} h_{0}^+$ vanishes due to the nature of anti-symmetry, although this term is allowed by $SU(2)_L\times U(1)_Y\times U(1)_{L_\mu - L_\tau}$.}.
The scalar potential of our model is given by
\begin{align}
V =& \mu_H^2 H^\dagger H + \lambda_H (H^\dagger H)^2 + \mu_\varphi^2 |\varphi|^2 + \lambda_\varphi |\varphi|^4 \nn \\
&+ M_{k^{++}}^2 k^{++} k^{--} + M_{h^+_{-1}}^2 h^+_{-1} h^-_{+1} + M_{h^+_{0}}^2 h^+_{0} h^-_{0} + M_{h^+_{+1}}^2 h^+_{+1} h^-_{-1}
 \nn \\
& + (\mu_{kh} k^{++} h^-_0 h^-_0 + \tilde \mu_{kh} k^{++} h^-_{-1} h^-_{+1}  + \mu_{\varphi h} \varphi h^+_{-1} h^-_{0} + \tilde \mu_{\varphi h} \varphi^* h^+_{+1} h^-_{0} + c.c.) \nn \\
& + (\lambda_{\varphi h k} \varphi k^{++} h^-_{-1} h_0^{-} + \tilde \lambda_{\varphi h k} \varphi^* k^{++} h^-_{+1} h^-_0 + c.c. ) + \lambda_{H k^{++}} (H^\dagger H) (k^{++} k^{--}) \nn \\
&  + \lambda_{H h_{-1}^+} (H^\dagger H)(h^+_{-1} h^-_{+1}) 
+ \lambda_{H h_{+1}^+} (H^\dagger H)(h^+_{+1} h^-_{-1}) + \lambda_{H h_{0}^+} (H^\dagger H)(h^+_{0} h^-_{0}) \nn \\
& + \lambda_{\varphi k^{++}} |\varphi|^2 (k^{++} k^{--}) + \lambda_{\varphi h_{-1}^+} |\varphi|^2 (h^+_{-1} h^-_{+1}) 
+ \lambda_{\varphi h_{+1}^+} |\varphi|^2 (h^+_{+1} h^-_{-1}) + \lambda_{\varphi h_{0}^+} |\varphi|^2 (h^+_{0} h^-_{0}) \nn \\
& + \lambda_{H\varphi} |\varphi|^2 (H^\dagger H)  + (\text{quartic terms for charged scalars }),
\end{align}
where we have omitted to write quartic terms containing only charged scalar fields and the coupling constants are assumed to be real for simplicity.

{\it CP-even scalar sector}:
After gauge symmetry breaking, we have two neutral physical scalar fields $\tilde{h}$ and $\varphi_R$.
The mass matrix for them is given by
\begin{equation}
L \supset \frac{1}{4} \begin{pmatrix} \tilde{h} \\ \varphi_R \end{pmatrix}^T
\begin{pmatrix} \lambda_H v_H^2 & \lambda_{H \varphi} v_H v_\varphi \\ \lambda_{H \varphi} v_H v_\varphi & \lambda_\varphi v_\varphi^2 \end{pmatrix}^T
\begin{pmatrix} \tilde{h} \\ \varphi_R \end{pmatrix}.
\end{equation}
Diagonalizing the mass matrix, we obtain mass eigenvalues and corresponding mass eigenstate $\{h, \phi\}$ such that
\begin{align}
& m_{h,\phi}^2 = \frac{\lambda_H v^2_H +\lambda_\varphi v_\varphi^2 }{4} \pm \frac{1}{4} \sqrt{\left( \lambda_H v^2_H -\lambda_\varphi v_\varphi^2 \right)^2 + 4 \lambda_{H \varphi}^2 v^2_H v_\varphi^2 }, 
 \\
& \begin{pmatrix} h \\ \phi \end{pmatrix} = \begin{pmatrix} \cos \theta & \sin \theta \\ - \sin \theta & \cos \theta \end{pmatrix} \begin{pmatrix} \tilde h \\  \varphi_R \end{pmatrix}, \quad
\tan 2 \theta = \frac{2 \lambda_{H \varphi} v_H v_\varphi}{\lambda_H v^2_H - \lambda_\varphi v_\varphi^2}.
\end{align}
In this paper we do not give further analysis for neutral scalar bosons where we assume mixing between $\phi$ and the SM Higgs is small to satisfy experimental constraints.
Phenomenology of the neutral scalar sector without charged Higgs bosons is given in ref.~\cite{Nomura:2018yej}.

{\it Charged scalar bosons}: Here we first consider mass of singly charged scalars where mass terms are given by
\begin{align}
L_{M_{h^+}} = & \left( M_{h^+_{-1}}^2  + \frac12 \lambda_{H h^+_{-1}} v^2_H + \frac12 \lambda_{\varphi h^+_{-1}} v_\varphi^2 \right) h^+_{-1} h^-_{+1} 
+ \left( M_{h^+_{0}}^2  + \frac12 \lambda_{H h^+_{0}} v^2_H + \frac12 \lambda_{\varphi h^+_{0}} v_\varphi^2 \right) h^+_{0} h^-_{0} \nn \\
& + \left( M_{h^+_{+1}}^2  + \frac12 \lambda_{H h^+_{+1}} v^2_H + \frac12 \lambda_{\varphi h^+_{+1}} v_\varphi^2 \right) h^+_{+1} h^-_{-1} + (m_A^2 h^+_{-1} h^-_{0} + m_B^2 h^+_{+1} h^-_{0} + c.c. ),
\end{align}
where we have defined $m_A^2 \equiv \mu_{\varphi h} v_\varphi/\sqrt2$ and $m_B^2 \equiv \tilde \mu_{\varphi h} v_\varphi/\sqrt2$.
In our scenario, we take $m_{A,B} \ll M_{h^+_{-1}, h^+_{0}, h^+_{+1}}$ so that mixing among singly charged scalars is negligibly small.
Moreover we obtain two-zero texture of neutrino mass matrix for small mixing case as we will see below.  
Therefore mass eigenstates of singly charged scalars are $\{ h^+_{-1}, h^+_{0}, h^+_{+1} \}$ whose masses are given by
\begin{equation}
m_{h^+_{-1,0,+1}}^2 \simeq M_{h^+_{-1,0,+1}}^2  + \frac12 \lambda_{H h^+_{-1,0,+1}} v^2 + \frac12 \lambda_{\varphi h^+_{-1,0,+1}} v_\varphi^2.
\end{equation}
On the other hand mass of doubly charged scalar $k^{\pm \pm}$ is given by
\begin{equation}
m_{k^{\pm \pm}}^2 = M^2_{k^{++}} + \frac12 \lambda_{H k^{++}} v^2 + \frac12 \lambda_{\varphi k^{++}} v_\varphi^2. 
\end{equation}

{\it Z' boson}: After $U(1)_{L_\mu - L_\tau}$ symmetry breaking, we have massive $Z'$ boson.
The mass of $Z'$ is given by
\begin{equation}
m_{Z'} = g' v_\varphi,
\end{equation}
where $g'$ is the $U(1)_{L_\mu - L_\tau}$ gauge coupling constant and we have ignored $U(1)$ kinetic mixing assuming it is negligibly small.
Gauge interactions of $Z'$ and the SM fermions are written as 
\begin{equation}
g' Z'_\nu (\bar L_\mu \gamma^\nu L_\mu - \bar L_\tau \gamma^\nu L_\tau + \bar \mu_R \gamma^\nu \mu_R - \bar \tau_R \gamma^\nu \tau_R).
\end{equation}

{\it Muon $g-2$}: In our model, $Z'$ and $h_{-1,0}^+$ bosons can provide contributions to muon $g-2$, $\Delta a_\mu$, at one-loop level. 
These contributions are given by~\cite{Babu:2002uu} 
\begin{align}
& \Delta a_\mu = \Delta a_\mu^{Z'} + \Delta a_\mu^{h^\pm} \nonumber \\
& \Delta a_\mu^{Z'} = \frac{g'^2}{8 \pi^2} \int_0^1 dx \frac{2 m_\mu^2 x^2 (1-x)}{x^2 m_\mu^2 + (1-x) m_{Z'}^2}, \quad
\Delta a_\mu^{h^\pm} \simeq - \frac{4 m_\mu^2}{96 \pi^2} \left( \frac{|f_{e\mu}|^2}{m_{h_{-1}^+}^2} + \frac{|f_{\mu \tau }|^2}{m_{h_{0}^+}^2} \right).
\end{align}
Note that charged scalar contribution provides negative sign contribution while $Z'$ gives positive one.

\begin{figure}[t] 
\begin{center}
\includegraphics[width=70mm]{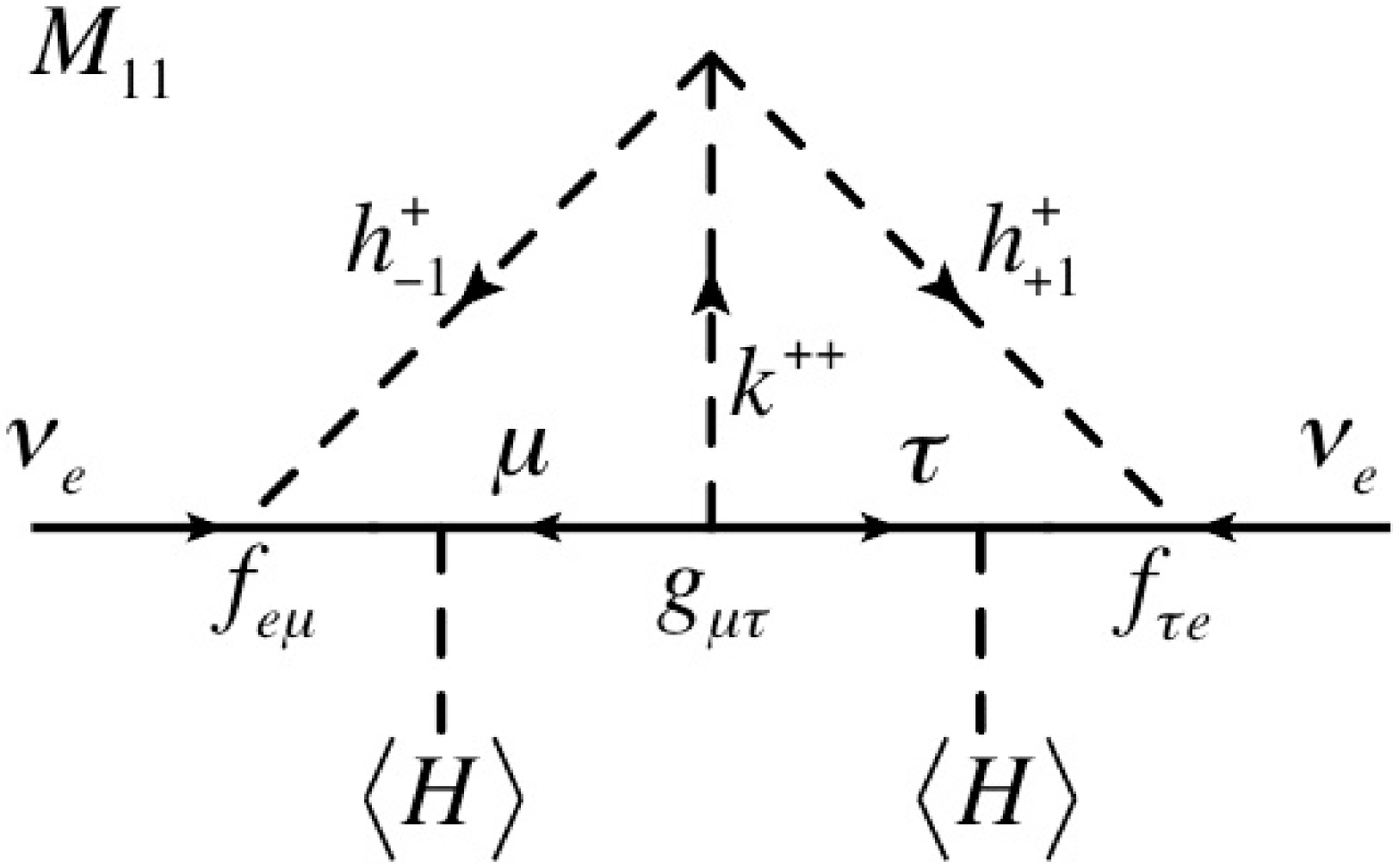} \qquad 
\includegraphics[width=70mm]{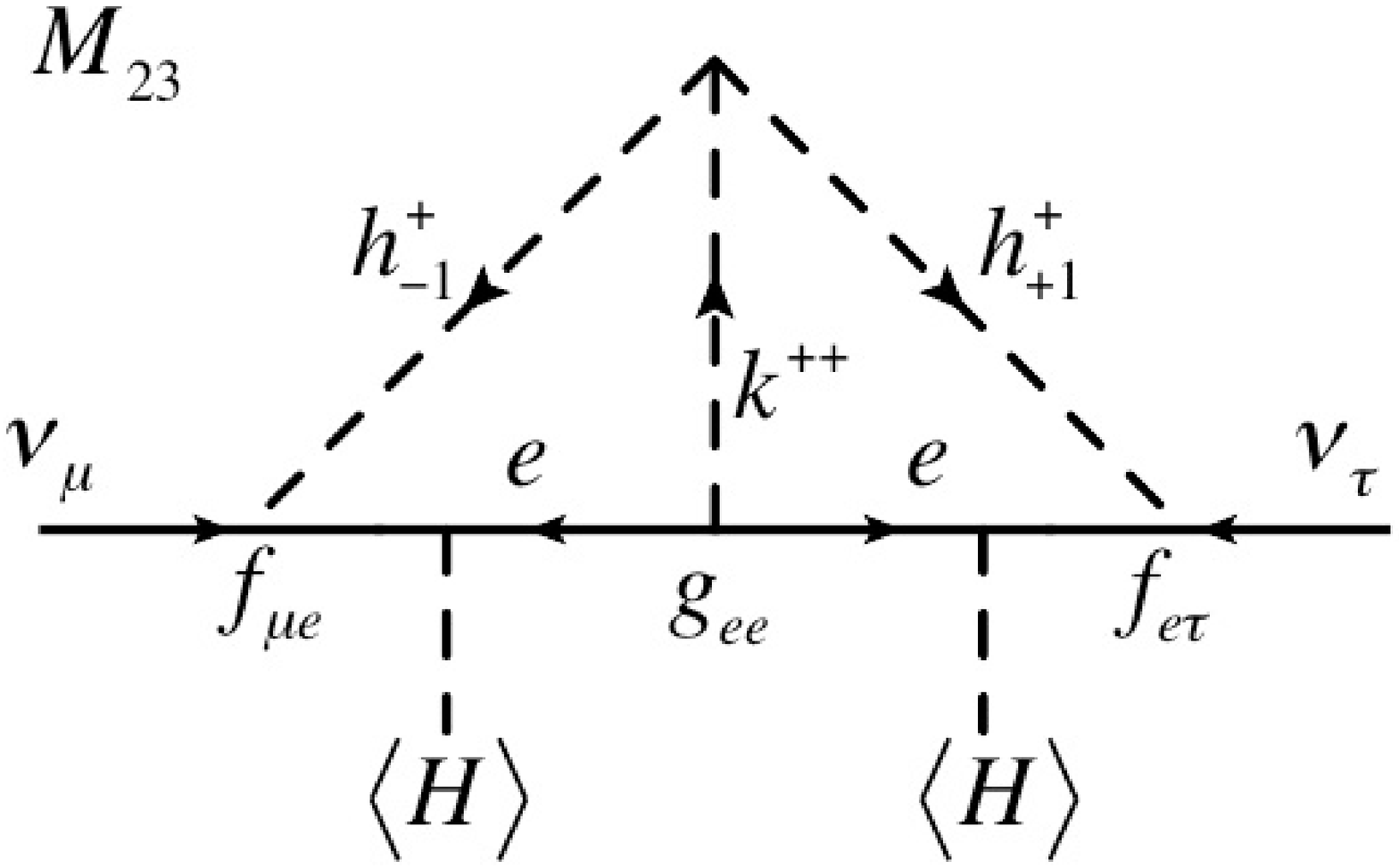}
\includegraphics[width=70mm]{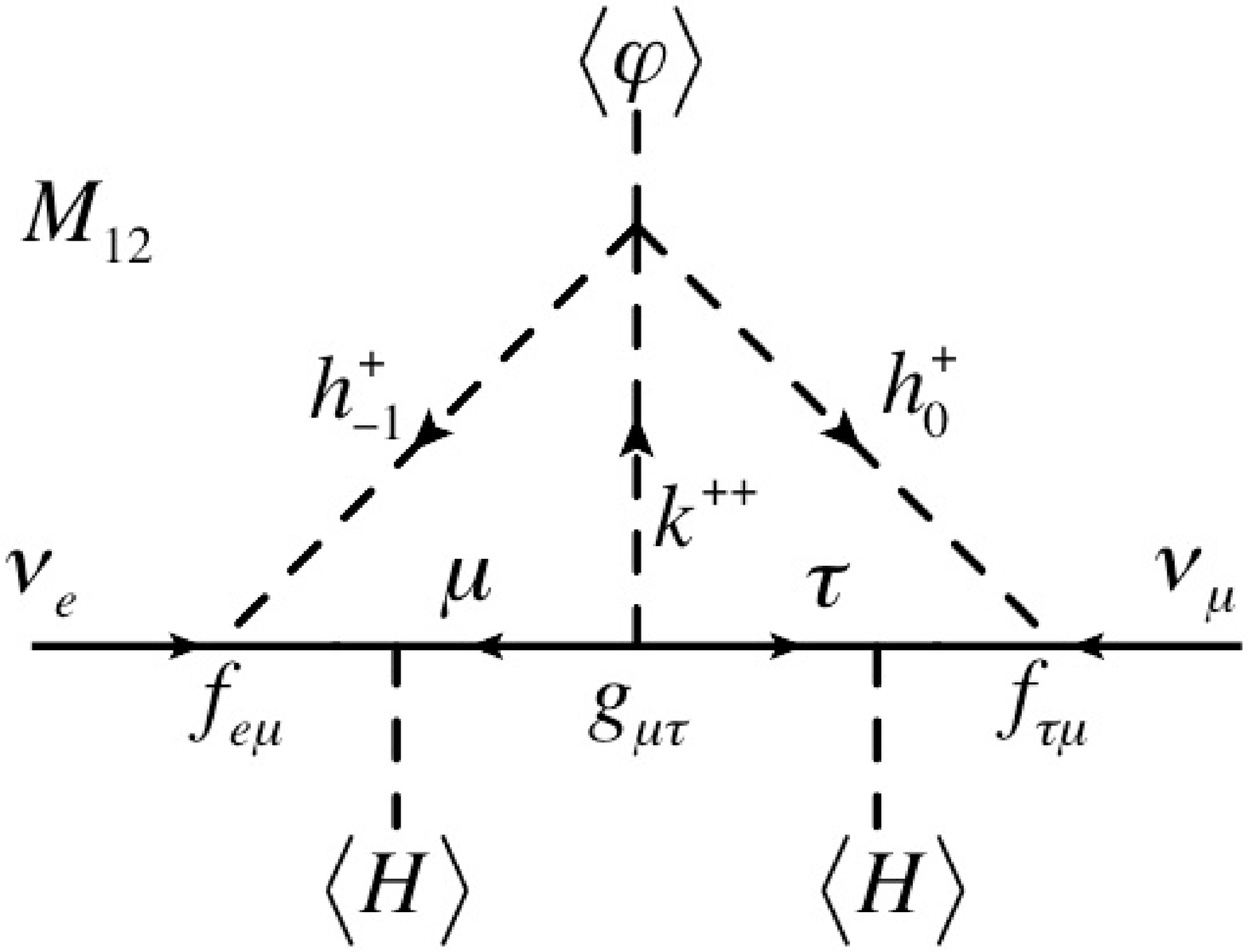} \qquad
\includegraphics[width=70mm]{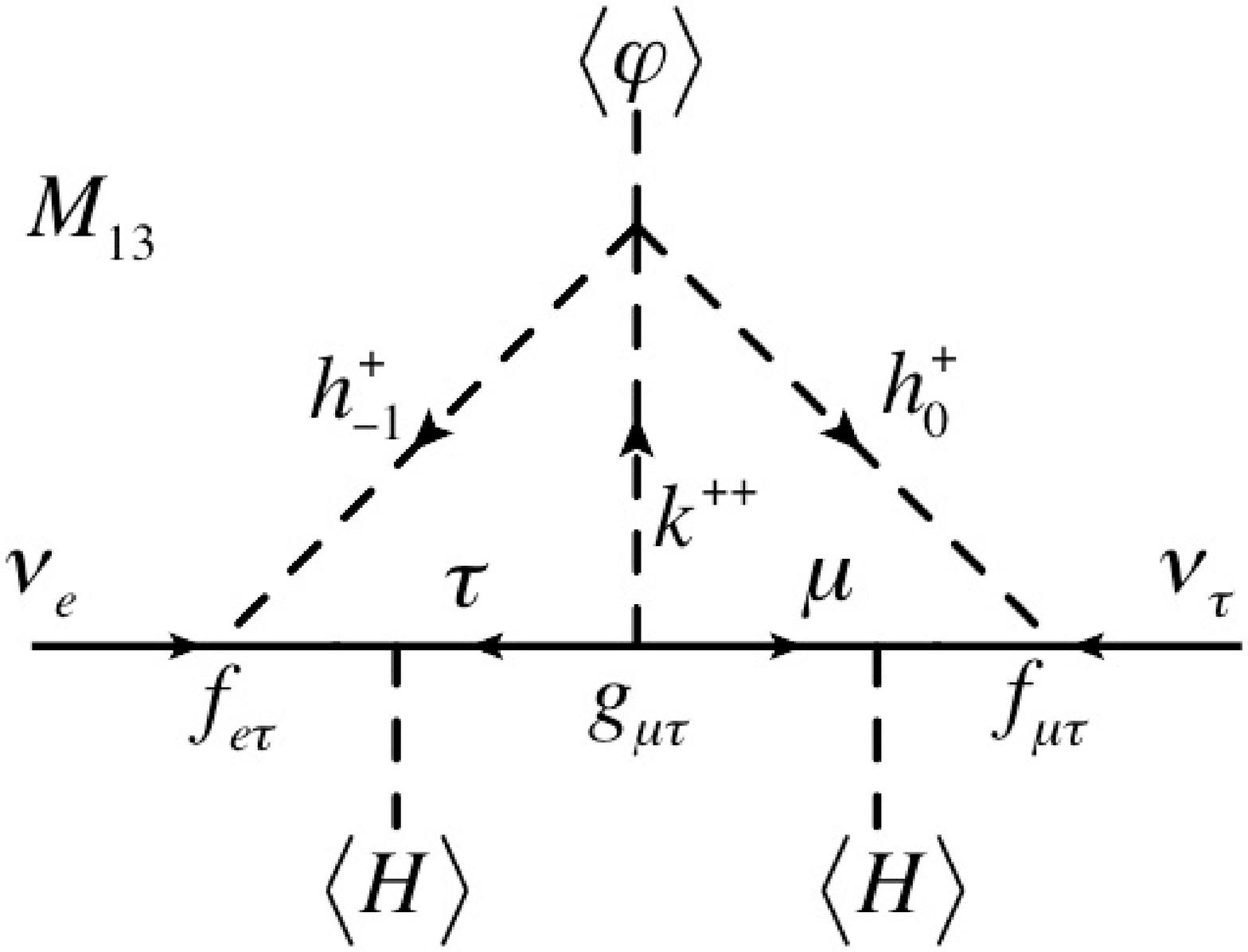}
\end{center}
\caption{Two loop diagrams generating neutrino mass matrix. 
\label{fig:diagram1} }
\end{figure}
{\it Neutrino mass matrix}: In our model, neutrino masses are generated at two loop level as the original Zee-Babu model~\cite{2-lp-zB}.
From the two loop diagrams in Fig.~\ref{fig:diagram1}, non-zero components of neutrino mass matrix are obtained such that
\begin{align}
M_{11} &= 8 \tilde \mu_{kh} f_{e \mu} m_\mu g_{\mu \tau}^* m_\tau f_{ \tau e} I \left( m_{h^+_{-1}}, m_{h^+_{+1}}, m_{k^{++}},m_\mu, m_\tau \right), \\
M_{12} &= 4\sqrt{2} \lambda_{\varphi h k} v_\varphi f_{e \mu} m_\mu g_{\mu \tau}^* m_\tau f_{ \tau \mu} I \left(  m_{h^+_{-1}}, m_{h^+_{0}}, m_{k^{++}}, m_\mu, m_\tau \right), \\
M_{13} &= 4\sqrt{2} \tilde \lambda_{\varphi h k} v_\varphi f_{e \tau} m_\tau g_{\mu \tau}^* m_\mu f_{\mu \tau} I \left( m_{h^+_{+1}}, m_{h^+_{0}}, m_{k^{++}}, m_\tau, m_\mu \right), \\
M_{23} &= 8 \tilde \mu_{kh} f_{ \mu e} m_e g_{ee}^* m_e f_{ e \tau } I \left( m_{h^+_{-1}}, m_{h^+_{+1}}, m_{k^{++}}, m_e, m_e \right), 
\end{align}
where the $I(m_1,m_2,m_3, m_{\ell_i}, m_{\ell_j})$ is the loop integral factor given by~\cite{McDonald:2003zj}
\begin{equation}
I(m_1,m_2,m_3, m_{\ell_i}, m_{\ell_j}) = \int \frac{d^4 p}{(2 \pi)^4} \int \frac{d^4 q}{(2 \pi)^4} \frac{1}{p^2 + m_1^2} \frac{1}{p^2 + m_{\ell_i}^2} \frac{1}{q^2 + m_2^2} \frac{1}{q^2 + m_{\ell_j}^2} \frac{1}{(q+p)^2 + m_3^2}.
\end{equation}
We thus obtain two-zero texture of the neutrino mass matrix in which $M_{33} \simeq M_{22} \simeq 0$ for small mixing among singly charged scalar bosons~\cite{Fritzsch:2011qv}.

Notice that the mass matrix can not be written as a product of $f$ and $g$ in contrast to the original Zee-Babu model,
and we can have three non-zero neutrino masses.
The loop integral factor is typically given by
\begin{equation}
I(m_1,m_2,m_3, m_{\ell_i}, m_{\ell_j}) \simeq \frac{C_I}{(4 \pi)^4} \frac{1}{M^2},
\label{eq:loopInt}
\end{equation}
where $C_I$ is $O(1)$ numerical factor~\cite{Babu:2002uu} and $M$ is the largest scalar mass inside the loop diagram.

\section{Phenomenology of the model}

In this section, we discuss phenomenology of our model.
Firstly we consider experimental constraints on the Yukawa couplings associated with charged scalar bosons.
The coupling $g_{ee}$ is constrained by the LEP data regarding $e^+e^- \to e^+ e^-$ scattering and the upper limit is written by~\cite{Nomura:2017abh}
\begin{equation}
|g_{ee}| \lesssim \frac{\sqrt{4 \pi} m_{k^{\pm \pm}}}{8.6 \ {\rm TeV}}.
\label{eq:LEPconst}
\end{equation} 
In addition, LFV process $\tau^\pm \to \mu^\mp e^\pm e^\pm$ gives the constraint~\cite{Akeroyd:2009nu, Nomura:2017abh} 
\begin{equation}
|g_{\mu \tau} g_{ee}^*| \lesssim 0.007 \left( \frac{m_{k^{\pm \pm}} }{\rm TeV} \right)^2,
\label{eq:LFVconst}
\end{equation} 
where the other LFV processes can be suppressed taking small mixing among singly charged scalar bosons.
Thus Yukawa couplings associated with singly charged scalars are less constrained in the small mixing case.

We then discuss requirements from neutrino mass matrix. 
Here we do not give detailed fitting to the neutrino oscillation data and use the result in ref.~\cite{Forero:2014bxa} where the mass matrix elements are required to be $\sim \mathcal{O}(10^{-11})$ GeV. 
The most stringent requirement comes from $M_{23}$ element since it is proportional to $m_e^2$ in our case. 
Using Eq.~(\ref{eq:loopInt}) for rough estimation, we obtain 
\begin{equation}
|M_{23}| \sim  3 \times 10^{-12} \ {\rm GeV} \times  \frac{|f_{\mu e}|}{10} \frac{|g_{ee}^*|}{0.4} \frac{|f_{e \tau}|}{10} C_I \frac{\tilde \mu_{kh}}{\rm TeV} \left( \frac{\rm TeV}{M} \right)^2 \, .
\end{equation}
Thus we should require $|f_{\mu e, e \tau}| \sim 10$, $\tilde \mu_{kh} \sim 10 M$ and $M$ to not be much larger than TeV scale to obtain $|M_{23}| \sim \mathcal{O}(10^{-11})$ GeV, taking into account the LEP constraint for $g_{ee}$ in Eq.~(\ref{eq:LEPconst}).
Moreover we find $g_{ee} \gg g_{\mu \tau}$ comparing $M_{11}$ and $M_{23}$ elements where the ratio of these couplings is roughly $m_\mu m_\tau/m_e^2$.
Then LFV constraint Eq.~(\ref{eq:LFVconst}) is easily satisfied as $g_{\mu \tau}$ should be much smaller than $g_{ee}$.
In general, we can easily get required values of the other matrix elements as we have sufficient free parameters.
Thus we focus on collider phenomenology for the doubly charged scalar boson associated with the coupling $g_{ee}$ in our following analysis.

Before we move to collider physics, we comment on the muon $g-2$ in our model. Since we require large magnitude of Yukawa coupling $|f_{\mu e}|$, $\Delta a_{\mu}^{h^+}$ get large negative contribution: 
\begin{equation}
\Delta a_\mu^{h^+} \sim - 4.7 \times 10^{-9} \left( \frac{|f_{\mu e}|}{10} \right)^2 \left( \frac{ {\rm TeV}}{m_{h^+_{0}}} \right)^2,
\end{equation}
where we have omitted contribution from $f_{\mu \tau}$ coupling assuming it is subdominant.
This negative contribution can still be compensated by the $Z'$ contribution.
For example, we obtain $\Delta a_\mu^{Z'} \simeq 6.5 \times 10^{-9}$ for $\{m_{Z'}, g' \}= \{9 \ {\rm MeV}, 0.0008\}$ which is allowed by the neutrino trident experiment~\cite{Altmannshofer:2014pba}.  
Thus it is possible to explain muon $g-2$, $\Delta a_\mu=(26.1\pm8.0)\times10^{-10}$~\cite{Hagiwara:2011af}, by the $Z'$ contribution even if we have negative contribution from singly charged scalar loop.~\footnote{Note here that $Z'$ does not affect the effective number of neutrino species and cosmology even if it is light since $Z'$ decouples from thermal bath earlier than neutrinos due to small gauge coupling and decays into neutrinos where the life time of $Z'$ with $\{m_{Z'}, g' \}= \{9 \ {\rm MeV}, 0.0008\}$ is of the order $10^{-15}$ second that is much shorter than the time of Big Bang nucleosynthesis.  }

\subsection{Doubly charged scalar production at hadron collider}

\begin{figure}[t]
\begin{center}
\includegraphics[width=70mm]{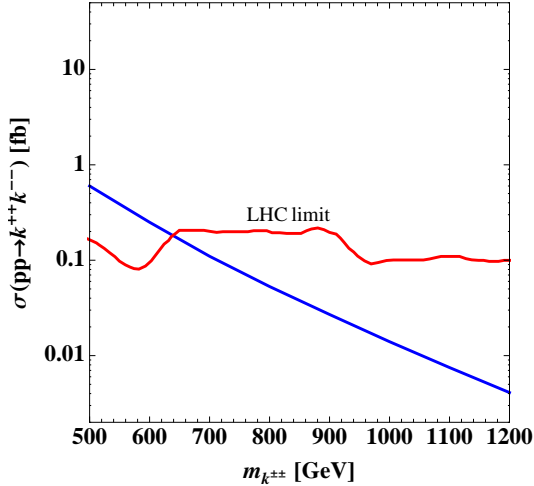} 
\caption{The blue curve show cross section for $pp \to k^{++} k^{--}$ as a function of $m_{k^{\pm \pm}}$ at the LHC 13 TeV.} 
  \label{fig:LHCcx}
\end{center}\end{figure}

The doubly charged scalar can be produced at the LHC via the process $pp \to Z/\gamma^* \to k^{++} k^{--}$ and dominantly decays into $e^\pm e^\pm$ due to the requirement for the Yukawa couplings discussed above.
Here we estimate the production cross section using {\tt CalcHEP}~\cite{Belyaev:2012qa} implementing relevant interactions with {\tt CTEQ6L} PDF~\cite{Nadolsky:2008zw}.
In Fig.~\ref{fig:LHCcx}, we show the cross section for LHC 13 TeV which is compared with the current LHC limit obtained from data of $\sim 36$ fb$^{-1}$ integrated luminosity~\cite{Aaboud:2017qph}.
Then the doubly charged scalar should be larger than $\sim 650$ GeV.
We can test heavier mass region with more integrated luminosity at the LHC or High luminosity LHC.
 If we take $m_{k^{\pm \pm}} = 1$ TeV the pair production cross section is around 0.014(0.019) fb for $\sqrt{s} = 13(14)$ TeV.
Thus the LHC data with 300(3000) fb$^{-1}$ integrated luminosity would provide few (few $\times 10$) number of events of $e^+e^-e^+e^-$ signature from decay of doubly charged scalar bosons.
Detailed simulation analysis is omitted here as it is beyond the scope of this paper.

\subsection{Testing doubly charged scalar Yukawa coupling at lepton collider}

Here we discuss test of doubly charged Yukawa coupling at the ILC with $\sqrt{s} = 250$ GeV.
Although doubly charged scalar can not be produced directly at the ILC, we can test the coupling observing deviation from the SM prediction for the $e^+ e^- \to e^+ e^-$ scattering.
The relevant effective interaction is written by
\begin{equation}
L_{eff} = \frac{g_{ee}^2}{2 m_{k^{++}}^2} (\bar e \gamma^\mu P_R e)(\bar e \gamma_\mu P_R e),
\end{equation}
where Fierz transformation is applied to get the operator.
We then apply the analysis in ref.~\cite{Nomura:2017abh} based on polarized initial state at the ILC, considering the process
\begin{align}
e^-(k_1,\sigma_1)  e^+(k_2,\sigma_2) \to \ell^-(k_3,\sigma_3)  \ell^+(k_4,\sigma_4),
\end{align}
where $k_i$ indicates 4-momentum of each particle and we explicitly show the helicities of initial- and final-state leptons $\sigma_{i} = \pm$.
Combining the SM and $k^{\pm \pm}$ contributions, total helicity amplitudes for the process of $e^-(\sigma_1)  e^+(\sigma_2) \to e^-(\sigma_3)  e^+(\sigma_4)$ denoted by ${\cal M}_{\sigma_i} = {\cal M}(\sigma_1  \sigma_2 \sigma_3 \sigma_4)$  are given by
\begin{align}
  & {\cal M}(+-+-) = -e^2\left(1+\cos\theta\right)
 \left[ 1 + \frac{s}{t} + c_R^2\left(\frac{s}{s_Z}+\frac{s}{t_Z}\right)
 + \frac{2 s}{\alpha (\Lambda_{RR}^e)^2}\right], \\ 
 & {\cal M}(-+-+) = -e^2\left(1+\cos\theta\right)
 \left[ 1 + \frac{s}{t} + c_L^2\left(\frac{s}{s_Z}+\frac{s}{t_Z}\right)
 \right], \\ 
 & {\cal M}(+--+) = {\cal M}(-++-) =
 e^2\left(1-\cos\theta\right)\left[1+c_Rc_L\frac{s}{s_Z}\right], \\
 & {\cal M}(++++) = {\cal M}(----) =
 2e^2\frac{s}{t}\left[1+c_Rc_L\frac{t}{t_Z}\right],
\end{align}
where $\Lambda_{RR}^e\equiv 4\pi m^2_{k^{\pm\pm}}/g^2_{ee}$, $t=(k_1-k_3)^2=(k_2-k_4)^2=-s(1-\cos\theta)/2$, $s=(k_1+k_2)^2=(k_3+k_4)^2$, 
$s_Z=s-m_Z^2+im_Z\Gamma_Z$, $t_Z=t-m_Z^2+im_Z\Gamma_Z$, 
$e^2=4\pi\alpha$ with $\alpha$ being the QED coupling constant,
$c_R=\tan\theta_W$, $c_L=-\cot2\theta_W$, and $\cos\theta$ is the scattering polar angle. 

The differential cross-section for purely-polarized initial-state $\sigma_{1,2} = \pm1$, is obtained using the amplitudes such as 
\begin{align}
 \frac{d\sigma_{\sigma_1\sigma_2}}{d\cos\theta} = \frac{1}{32\pi s}
 \sum_{\sigma_3,\sigma_4} \left|{\cal M}_{\{\sigma_i\}}\right|^2.
\end{align}
Then the partially-polarized differential cross section is defined as
\begin{align}
\frac{d \sigma (P_{e^-}, P_{e^+})}{d \cos \theta} = \sum_{\sigma_{e^-}, \sigma_{e^+} = \pm} \frac{1+ \sigma_{e^-} P_{e^-}}{2} \frac{1 +\sigma_{e^+} P_{e^-}}{2} \frac{d \sigma_{\sigma_{e^-} \sigma_{e^+}}}{d \cos \theta},
\end{align}
where $P_{e^-(e^+)}$ is the degree of polarization for the electron(positron) beam and we sum up the helicity of final states. 
We then apply the following two cases as realistic values at the ILC for polarized cross sections $\sigma_{L,R}$:
\begin{equation}
\frac{d \sigma_{R}}{d \cos \theta} = \frac{d \sigma (0.8,-0.3)}{d \cos \theta}, \quad \frac{d \sigma_{L}}{d \cos \theta} = \frac{d \sigma (-0.8,0.3)}{d \cos \theta}.
\end{equation}
Applying the polarized cross sections, we study the sensitivity to $k^{\pm \pm}$ boson in $e^+ e^- \to e^+ e^-$ scattering 
via the measurement of a forward-backward asymmetry at the ILC, which is given by 
\begin{align}
& A_{FB} (\sigma_{L,R}) = \frac{N_F (\sigma_{L,R}) - N_B (\sigma_{L,R})}{N_F (\sigma_{L,R}) + N_B (\sigma_{L,R})}, \nonumber \\
 & N_{F(B)} (\sigma_{L,R}) = L \int_{0(-0.5)}^{0.5 (0)} d \cos \theta \frac{d \sigma_{L,R}}{d \cos \theta},
\end{align}
where $L$ is an integrated luminosity, and a bound of integral $\pm 0.5$ is chosen to maximize the sensitivity.
Then the forward-backward asymmetry is estimated for cases with only the SM gauge boson contributions, and with both SM and $k^{\pm \pm}$ boson contributions,
in order to explore the sensitivity to $k^{\pm \pm}$ interaction.
We thus obtain $N_{F(B)}^{SM}(\sigma_{L,R})$ and $A_{FB}^{SM}(\sigma_{L,R})$ for the former case, and $N_{F(B)}^{SM+k^{\pm \pm}}(\sigma_{L,R})$ and $A_{FB}^{SM+k^{\pm \pm}}(\sigma_{L,R})$ for  the latter case. 
Finally the sensitivity to $k^{\pm \pm}$ interaction is estimated by 
\begin{equation}
\Delta A_{FB}(\sigma_{L,R}) = |A_{FB}^{SM+k^{\pm \pm}} (\sigma_{L,R})- A_{FB}^{SM}(\sigma_{L,R})|.
\label{eq:delAFB}
\end{equation}
Then we compare this quantity with a statistical error of the asymmetry which is given by assuming only SM contribution:
\begin{equation}
\delta_{A_{FB}}^{SM}(\sigma_{L,R}) = \sqrt{\frac{1-(A_{FB}^{SM}(\sigma_{L,R}))^2}{N_F^{SM}(\sigma_{L,R})+N_B^{SM}(\sigma_{L,R})}}.
\label{eq:SMerror}
\end{equation}

\begin{figure}[t]
\begin{center}
\includegraphics[width=70mm]{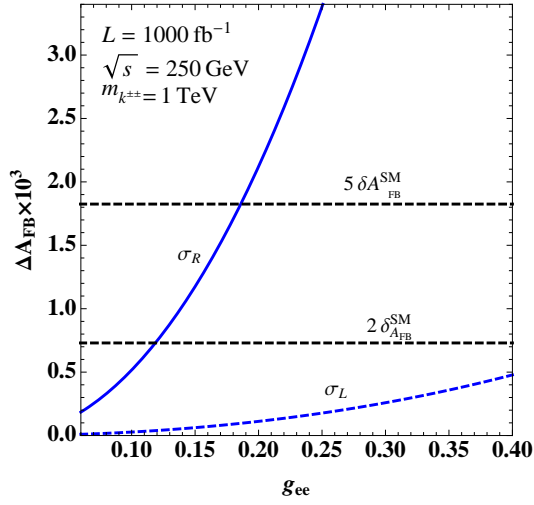} 
\caption{The blue solid(dashed) curve shows $\Delta A_{FB}$ defined as Eq.~(\ref{eq:delAFB}) as a function of $g_{ee}$ for $\sigma_{R(L)}$. The statistical error in the SM, $\delta_{A_{\rm FB}}^{\rm SM}$ given by Eq.~(\ref{eq:SMerror}), is estimated to be $\sim 0.36 \times 10^{-3}$ both $\sigma_R$ and $\sigma_L$.} 
  \label{fig:AFBe}
\end{center}\end{figure}

In Fig.~\ref{fig:AFBe}, we show $\Delta A_{FB}(\sigma_R)$ and $\Delta A_{FB}(\sigma_L)$ by solid and dashed curves respectively as a function of $g_{ee}$
where we apply integrated luminosity of 1000 fb$^{-1}$ as a reference value. 
The curves are compared with the values $5 \delta_{A_{FB}}^{SM}$ and $2 \delta_{A_{FB}}^{SM}$ which are respectively given by $\sim 7.2 \times 10^{-3}$ and $\sim 1.8 \times 10^{-3}$.
Thus we find that $g_{ee} \gtrsim 0.12$ can be tested with the integrated luminosity of 1000 fb$^{-1}$ with 2 $\sigma$ level and 5 $\sigma$ significance can be obtained for $g_{ee} \gtrsim 0.18$ 
for polarized cross section $\sigma_R$. On the other hand, $\Delta A_{FB}(\sigma_L)$ is much smaller than that for $\sigma_R$. Therefore comparing two polarized cross section cases we can test right-handed property of the Yukawa coupling. Moreover applying larger integrated luminosity as 2000 fb$^{-1}$ we can test the whole region of $g_{ee} > 0.1$ which is preferred by the neutrino mass matrix.  

\section{Conclusion}
We have proposed a model providing the neutrino mass and mixing by extending the Zee-Babu model imposing gauged $U(1)_{L_\mu- L_\tau}$ symmetry and introducing several charged scalar fields.
Due to restricted Yukawa couplings resulting from the additional symmetry, we have found a predictive neutrino texture in a simple hypothesis in which mixings among singly-charged scalar bosons are negligibly tiny. 
In addition, LFV constrains are much relaxed compared to the original model in the small mixing scenario. 
 The structure of neutrino mass matrix also constrains the relative values of the Yukawa couplings associated with doubly-charged scalar and the masses of charged scalars are preferred to be around TeV scale. 
We have also shown that anomalous muon magnetic dipole moment can be explained by $Z'$ boson contribution even if singly charged scalars give opposite sign contributions.

Then we have focussed on analyzing phenomenology of doubly-charged scalar boson in the LHC and the future ILC experiments.
The doubly-charged scalars can be produced in pair at the LHC if its mass is around TeV and $e^+e^-e^+e^-$ signature will be dominant since electron Yukawa coupling is required to be the largest from our neutrino mass matrix. With sufficient integrated luminosity, doubly-charged Higgs can be discovered at the LHC and the high luminosity LHC experiments.
On the other hand effect of doubly-charged scalar can be explored by measuring $e^+e^- \to e^+ e^-$ scattering at the ILC although direct production is not kinematically allowed.
Considering forward backward asymmetry in the process with polarized beam, we have estimated the sensitivity to the doubly-charged scalar interaction at the ILC with $\sqrt{s}=250$ GeV.
We then find that a Yukawa coupling $g_{ee}$ up to $\sim$0.1 can be tested for $m_{k^{\pm\pm}}\approx 1$ TeV where those parameter region are preferred to get viable neutrino mass matrix.
Therefore doubly-charged scalar interactions in our model can be tested in future collider experiments in which the LHC and the ILC will give complementary results.

\section*{Acknowledgments}
\vspace{0.5cm}
H.~O.\ is sincerely grateful for all the KIAS members. 

\end{document}